\documentclass
[preprint,fleqn,a4paper,11pt,amsfonts,amssymb,nobibnotes,prd,titlepage,superscriptaddress,nofootinbib]{revtex4}%
\usepackage[dvips]{graphicx}
\usepackage{amsfonts}
\usepackage{amssymb}
\usepackage{amsmath}%
\providecommand{\U}[1]{\protect\rule{.1in}{.1in}}

\begin{document}
\preprint{KEK Preprint 2015-48}
\title{Schwinger-Dyson Study for Walking/Conformal Dynamics with IR Cutoffs}
\author{Kohtaroh Miura}
\email{Kohtaroh.Miura@cpt.univ-mrs.fr}
\affiliation{Kobayashi-Maskawa Institute for the Origin of Particles and the Universe
(KMI), Nagoya University, Nagoya, 464-8602, Japan}
\affiliation{Centre de Physique Theorique (CPT), Aix-Marseille University, Campus de
Luminy, Case 907, 163 Avenue de Luminy, 13288 Marseille cedex 9, \ France}
\author{Kei-ichi Nagai}
\email{keiichi.nagai@kmi.nagoya-u.ac.jp}
\affiliation{Kobayashi-Maskawa Institute for the Origin of Particles and the Universe
(KMI), Nagoya University, Nagoya, 464-8602, Japan}
\author{Akihiro Shibata}
\email{akihiro.shibata@kek.jp}
\affiliation{Computing Research Center, High Energy Accelerator Research Organization(KEK),
Tsukuba 305-0801, Japan}
\begin{abstract}
Motivated by recent progress on many flavor QCD on a lattice, we investigate
conformal/walking dynamics by using Schwinger-Dyson (SD) equation within an
improved ladder approximation for two-loop running coupling. By numerically
solving the SD equation, we obtain a pole mass $m_{p}$, pion decay constant
$f_{\pi}$, and investigate the chiral symmetry breaking and mass anomalous
dimension $\gamma_{m}$ in the presence of IR cutoffs $\Lambda_{\mathrm{IR}}$.
We find that the chiral symmetry breaking is suppressed \ if IR cutoff
$\Lambda_{\mathrm{IR}}$ becomes larger than the critical \ value near the
dynamical mass ($\Lambda_{\mathrm{IR}}$ $\simeq m_{D}$) In the conformal phase
the $\gamma_{m}$ is strongly suppressed by IR cutoffs for $\Lambda
_{\mathrm{IR}}$ $\simeq m_{p}$. We, then, obtain finite size hyperscaling
(FSS) relation by adapting a linearized approximation for the SD equation, and
examine the $\gamma_{m}$ The results offer valuable insight and suggestion for
analyses in lattice gauge theories.

\end{abstract}
\maketitle

\section{Introduction}

When a fermion multiplicity (flavor) exceeds a critical value $N_{f}%
=N_{f}^{\ast}$ in non-Abelian gauge theories, a quantum phase transition from
a chiral broken phase to the conformal window is anticipated to take place. In
the conformal window, strong coupling gauge theories are asymptotically-free
in ultra violet (UV) region, while dominated by a non-trivial infrared fixed
point (IRFP) in infrared (IR) region. The IRFP-conformal dynamics can be
probed via a response to a small fermion mass ($m_{0}$) perturbation, which
gives rise to a bound state with a mass gap ($M_{\mathrm{H}}$) satisfying the
conformal hyperscaling relation with a mass anomalous dimension $\gamma_{m}$.
In the chiral broken phase near to the conformal window $N_{f}\lesssim
N_{f}^{\ast}$, it is advocated that the system shows an approximate
conformality (walking) with a ``would-be'' $\gamma_{m}$, which is of great
interest in physics beyond the Standard Model \cite{review-composite-higgs}.
In recent lattice studies, the precise determination of $\gamma_{m}$ in many
flavor QCD is reported, while there exist tensions among the values
\cite{review-gamma}. The discrepancy presumably originates to scale violations
by UV/IR cutoffs (lattice spacing and finite box size) as well as a probe
fermion mass. In this proceedings, we focus on the IR cutoff effects by using
Schwinger-Dyson (SD) equation.

\section{Schwinger-Dyson equation analysis}

We consider the SD equation for the fermion propagator
$iS_{\mathrm{F}}(q)\equiv\bigl[{}\hspace{-1.0mm}\not \hspace {-1.0mm}%
{p}\hspace{--1.0mm} - \Sigma(p^{2})\bigr]^{-1}$ within improved ladder
approximation in the Landau gauge. After integrating over angular degrees of
freedom in a momentum space, the SD equation is expressed as
\begin{align}
\Sigma(p^{2}) = m_{0}+\int_{\Lambda_{\mathrm{IR}}}^{\Lambda_{\mathrm{UV}}}
dq^{2} \frac{3C_{2}[F]\alpha(p^{2}+q^{2})}{4\pi} \left[  \frac{q^{2}}{p^{2}%
}\theta_{p^{2} - q^{2}}+\theta_{q^{2}-p^{2}}\right]  \frac{\Sigma(q^{2}%
)}{q^{2}+\Sigma^{2}(q^{2})}, \label{SDeq}%
\end{align}
where $m_{0}$, $\theta_{p^{2}}$, $\alpha(\mu) = g^{2}(\mu)/(4\pi^{2})$, and
$C_{2}[F]$ represent a bare fermion mass, step function, running coupling
constant in the two-loop perturbation theory, and quadratic Casimir operator,
respectively.
The UV/IR cutoffs $\Lambda_{\mathrm{UV/IR}}$ are introduced in the momentum
space integral. For a given fermion mass $m_{0}$ and in the presence of
$\Lambda_{\mathrm{UV/IR}}$, we numerically solve the SD equation (\ref{SDeq})
and evaluate the physical pole mass $m_{p}$ and pion decay constant $f_{\pi}$
as
\begin{equation}
m_{p}\equiv\Sigma(p^{2}=m_{p})\ ,\quad f_{\pi}^{2}=\frac{N_{c}}{4\pi^{2}}
\int_{0}^{\Lambda_{\mathrm{UV}}^{2}}dz\ z\ \frac{\bigl(1-\frac{1}{4}z\frac{d}
{dz}\bigr)\Sigma^{2}(z)}{\bigl(z+\Sigma^{2}(z)\bigr)^{2}}\ . \label{eq:mp_fpi}%
\end{equation}
We take account of full momentum dependences of the two-loop running coupling
without recourse to the usual step-function type approximation for the
coupling. We then perform the hyperscaling fit analyses for the obtained
numerical data similarly to analyses in lattice gauge theories. Moreover, we
derive the SD-based finite-size hyperscaling formula (FSS), which allows us to
handle the IR cutoff artifacts, even in the case of $\Lambda_{\mathrm{IR}}
\gtrsim m_{p},f_{\pi}$. These are the advantages to the previous work
\cite{Aoki:2012ve}.

\begin{figure}[tb]
\includegraphics[width=5.1cm]{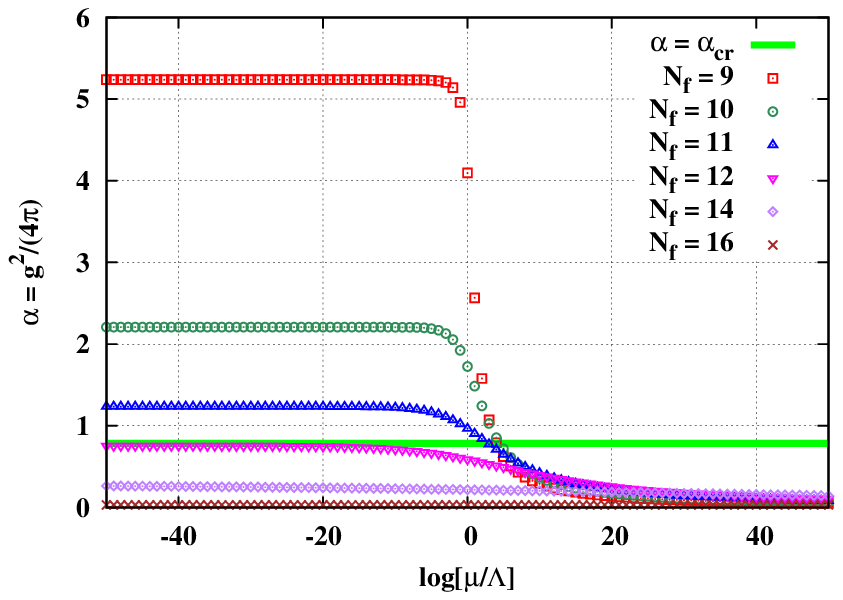}
\includegraphics[width=5.1cm]{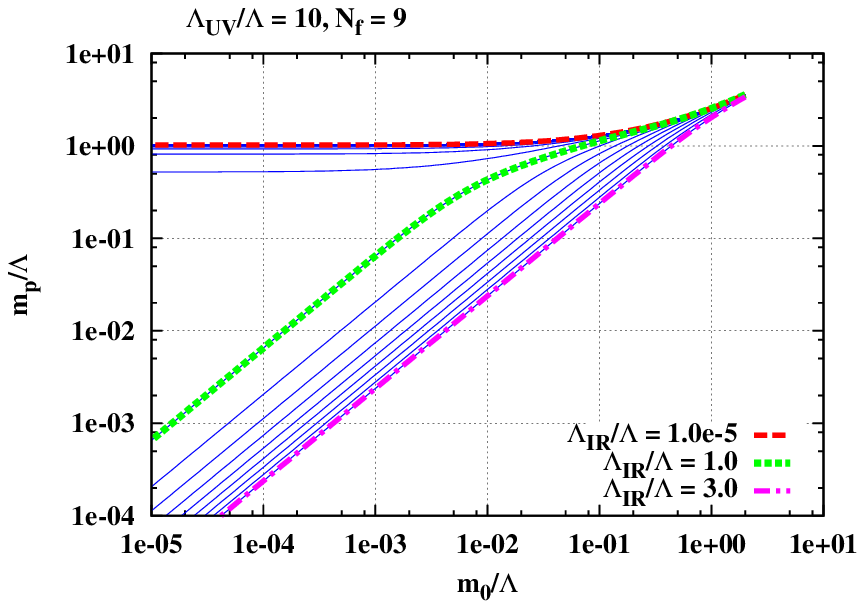}
\includegraphics[width=5.1cm]{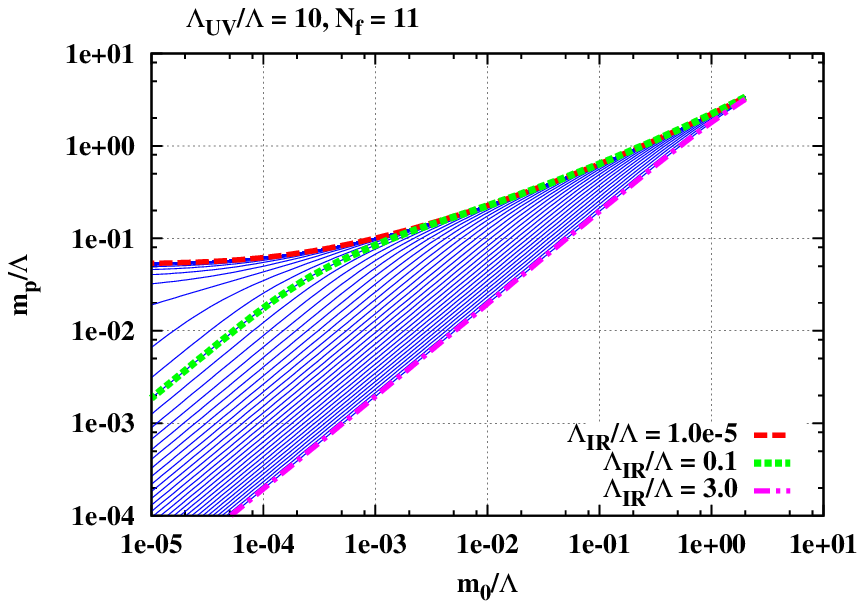} \caption{ Left: The
running coupling for the various fermion multiplicity $9\leq N_{f}\leq16$ in
the color SU($N_{c}=3$) gauge theory. Middle and Right: Pole masses $m_{p}$ as
a function of a bare fermion mass $m_{f}$ for various IR cutoffs
($\Lambda_{\mathrm{IR}}/\Lambda\in\lbrack10^{-5},3.0]$, blue-thin-solid lines)
with the UV cutoff $\Lambda_{\mathrm{UV}}/\Lambda=10$. in the chiral broken
phase: (Middle panel) $N_{f}=9$, ( Right) $N_{f}=11$.}%
\label{fig:SDeq-sol}%
\end{figure}


\section{Chiral broken and conformal phases with IR cutoff}

First, we investigate the chiral broken phase in the presence of IR cutoffs.
The middle and right panels of Fig.~\ref{fig:SDeq-sol} show the pole mass
$m_{p}$ against $m_{0}$ for $N_{f} = 9$ and $11$, respectively. Each
blue-thin-solid line represents a result with a different IR cutoff
$\Lambda_{\mathrm{IR}}/\Lambda$ \footnote{The scale $\Lambda$ denotes the
intrinsic scale which is renormalization invariant and associated with the
onset of non-perturbative dynamics \cite{Appelquist:1998rb}.}. For a small
$\Lambda_{\mathrm{IR}}/\Lambda$, the pole masses remain finite at small
$m_{0}$, indicating the spontaneous chiral symmetry breaking.
We find a critical value of the IR cutoff above which the dynamical fermion
mass $m_{D}$ is strongly suppressed as shown in the figures:
\begin{equation}
\Lambda_{\mathrm{IR}}^{\ast}=m_{p}(\Lambda_{\mathrm{IR}}=0,m_{0}=0)=:m_{D} .
\label{eq:IRcritical}%
\end{equation}
Thus, a \textit{Fake Conformality} results from a large $\Lambda_{\mathrm{IR}%
}$. As a system becomes closer to the conformal window ($N_{f} = 9 \to11$),
the fake conformality appears for a smaller $\Lambda_{\mathrm{IR}}$. This
gives a caveat for lattice works studying the border region of chiral broken
and conformal phases.

Next, we investigate a conformal phase with IR cutoffs by using the SD results
for $N_{f} = 12$. The mass anomalous dimension $\gamma_{m}$ is obtained by the
scaling property of $m_{p}$ ($f_{\pi}$) against $m_{0}$. The left panel of
Fig.~\ref{fig:FSS} shows a logarithm plot of the pole mass $m_{p}$ as a
function of fermion bare masses $m_{0}$ for various IR cutoffs $\Lambda
_{\mathrm{IR}}$. For the smallest IR cutoff $\Lambda_{\mathrm{IR}}=10^{-8}$
(red-squares), the data points align on a straight line, therefore, the
hyperscaling relation $m_{p}=Cm_{0}^{1/(1+\gamma_{m})}$ is satisfied. For the
fit range of $m_{0}\in[0.001,0.1]$ (non-shaded region in the figure), we
obtain $\gamma_{m}=0.56$. While, for larger IR cutoffs, the data points are on
a polygonal line with a bend around $\Lambda_{\mathrm{IR}}\simeq m_{p}$. As
seen in the figure, the fit range $m_{0}\in\lbrack0.001,0.1]$ starts affected
by the bend down with increasing $\Lambda_{\mathrm{IR}}$. As a result, the fit
with the ansatz $m_{p}=Cm_{0}^{1/(1+\gamma_{m})}$ results in a $\gamma_{m}$
strongly suppressed, as shown in the middle panel of Fig.~\ref{fig:FSS}. When
the suppression sets in, the $\gamma_{m}$ evaluated by $m_{p}$ (red-squares)
and $f_{\pi}$ (blue-circles) get impaired, and thus the universality of
$\gamma_{m}$ does not hold any more.

\section{Finite size hyperscaling (FSS) and mass anomalous dimension}

We shall now derive a SD-based FSS which allows us to handle the IR cutoff
artifacts explained above. To this end, we further approximate the SD equation
(\ref{SDeq}) with a linearized fermion propagator and $\alpha(\mu)\to
\alpha_{*}\theta_{\Lambda-\mu}$ where $\alpha_{*}$ denotes the running
coupling at IRFP, and obtain the analytic expression of the renormalization
factor $Z_{M}^{-1}:=m_{0}/m_{p}$. The $Z_{M}$ is a function of dimensionless
variables, $\hat{m}_{p}:=m_{p}/\Lambda$, $\hat{m}_{0}:=m_{0}/\Lambda$,
$\hat{\Lambda}:=\Lambda_{\mathrm{IR}}/\Lambda$, and $\gamma=1-\sqrt
{1-\alpha_{\ast}/\alpha_{\mathrm{cr}}}$, and the analytic expression allows us
to derive the SD-based FSS formula \cite{MNS2014}:
\begin{align}
&  \hat{m}_{p}/\hat{\Lambda}_{\mathrm{IR}} = C_{\mathrm{X}}\cdot X \ ,\quad X
\equiv\hat{m}_{0}^{1/(1+\gamma_{m})}/\hat{\Lambda}_{\mathrm{IR}}
\ ,\quad\Lambda_{\mathrm{IR}}\ll m_{p}\ll\Lambda\ ,\label{eq:fss0}\\
&  \hat{m}_{p}/\hat{\Lambda}_{\mathrm{IR}} = C_{\mathrm{Y}}\cdot Y \ ,\quad
Y\equiv\hat{m}_{0}/\hat{\Lambda}_{\mathrm{IR}}^{1+\gamma_{m}} \ ,\quad m_{p}
\ll\Lambda_{\mathrm{IR}} \ll\Lambda\ . \label{eq:fssir}%
\end{align}
The equation (\ref{eq:fss0}) is same as what dictated by the renormalization
group equation \cite{DelDebbio:2010ze}. Remarkably, the SD predicts another
FSS formula (\ref{eq:fssir}) which is characterized by the scaling variable
$Y$ rather than $X$ for larger $\Lambda_{\mathrm{IR}}$.

\begin{figure}[th]
\includegraphics[width=5.1cm]{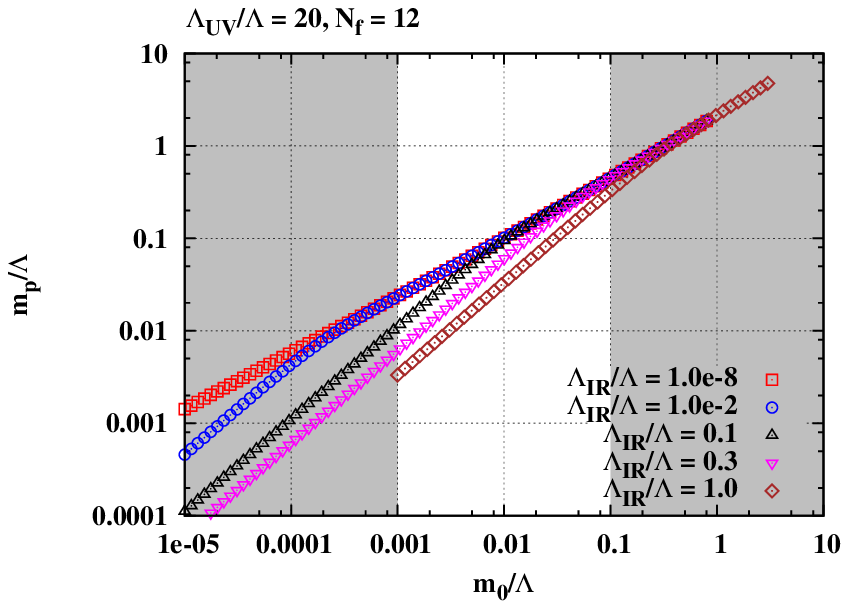}
\includegraphics[width=5.1cm]{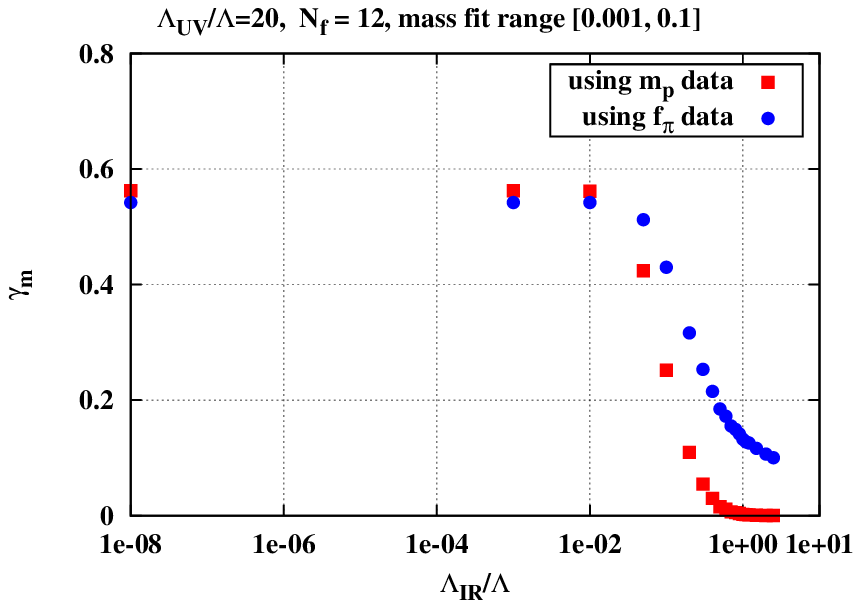}
\includegraphics[width=5.1cm]{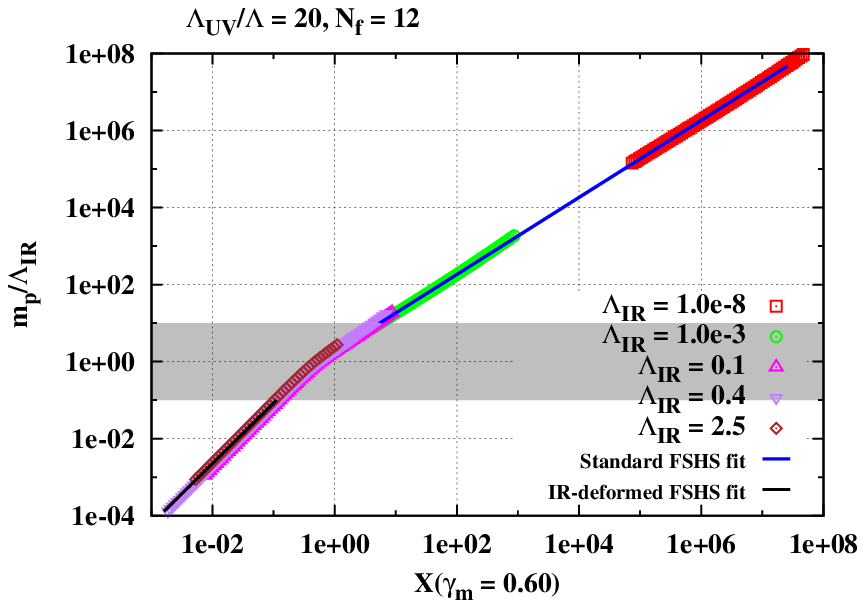} \caption{ Left:
The pole mass $\hat{m}_{p}=m_{p}/\Lambda$ vs the fermion mass $\hat{m}%
_{0}=m_{0}/\Lambda$ for various IR cutoffs $\hat{\Lambda}_{\mathrm{IR}} =
\Lambda_{\mathrm{IR}}/\Lambda$ in $N_{f} = 12$. Middle: The mass anomalous
dimension $\gamma_{m}$ obtained by fitting the non-shaded region data of the
left panel with the conformal ansatz. Right: The pole mass $\hat{m}_{p}$ vs $X
= \hat{m}_{0}^{1/(1+\gamma_{m})}/\hat{\Lambda}_{\mathrm{IR}}|_{\gamma_{m} =
0.60}$ in the conformal phase $N_{f} = 12$. The data upper and lower than the
shaded region are used for the fits, giving the blue- and black-solid lines. }%
\label{fig:FSS}%
\end{figure}

We apply the SD-based FSS (\ref{eq:fss0}) and (\ref{eq:fssir}) for the data
obtained by solving Eq.~(\ref{SDeq}) numerically. Right panel of
Fig.~\ref{fig:FSS} shows the results. The various symbols (colors) show data
with different IR cutoffs. For the case of $\Lambda_{\mathrm{IR}}\ll m_{p}%
\ll\Lambda$, we adopt the ansatz (\ref{eq:fss0}). The fit works well
(blue-solid line) and we obtain $\gamma_{m}\simeq0.60 =:\gamma_{1}$ which is
fairly consistent to $\gamma_{m}=0.56$ obtained in the previous section for
the smallest IR cutoff $\hat{\Lambda}_{\mathrm{IR}}=10^{-8}$. We find the
alignment of the data points, indicating the universal nature of $\gamma
_{m}=0.60$. For the case of $m_{p}\ll\Lambda_{\mathrm{IR}}\ll\Lambda$, we
adopt the ansatz (\ref{eq:fssir}). The fit works well (black-solid line) and
gives $\gamma_{m}\simeq0.56 =: \gamma_{2}$, which is somewhat smaller than
$\gamma_{1}=0.60$ but the strong suppression has disappeared. Thus, the right
FSS formula in the right place recovers the approximate universality
$\gamma_{1}\simeq\gamma_{2}$, or equivalently, the approximate data alignments
in the whole mass region including both $\Lambda_{\mathrm{IR}}\ll m_{p}%
\ll\Lambda$ and $m_{p}\ll\Lambda_{\mathrm{IR}}\ll\Lambda$. Then, two scaling
variables, $X(\gamma_{1})$ and $Y(\gamma_{2})$, are responsible for two slopes
of the alignments. In lattice studies, indeed, the formula (\ref{eq:fss0}) is
widely used in lattice study, but the formula (\ref{eq:fssir}) is rarely used
\cite{Degrand}. However, these result are of the case without paying attention
to the scope of application, and opposed to each other \cite{review-gamma}.

\begin{acknowledgments}
This work is supported by the JSPS Grant-in-Aid for Scientific Research (S)
No.22224003. We thank Robert Shrock and Seyong Kim for fruitful discussions
during the conference.
\end{acknowledgments}



\end{document}